\documentclass[apj]{emulateapj}

\usepackage{graphicx,times}
\usepackage{subfigure}
\newcommand{\be}{\begin{equation}}
\usepackage{threeparttable}
\usepackage{booktabs}
\newcommand{\ee}{\end{equation}}
\newcommand{\bea}{\begin{eqnarray}}
\newcommand{\eea}{\end{eqnarray}}

\usepackage{enumerate}
\usepackage{amsmath}
\usepackage{cases}
\usepackage{longtable}
\usepackage{hyperref}
\usepackage{epstopdf}
\usepackage{amsmath,bm}
\usepackage{amssymb}
\usepackage{natbib}
\usepackage{morefloats}
\usepackage{multirow}
\usepackage{array}
\usepackage{verbatim}

\begin{document}

\title{Probing the intergalactic turbulence with fast radio bursts}

\author{Siyao Xu\altaffilmark{1} and Bing Zhang \altaffilmark{2}}

\altaffiltext{1}{Department of Astronomy, University of Wisconsin, 475 North Charter Street, Madison, WI 53706, USA; 
Hubble Fellow, sxu93@wisc.edu}
\altaffiltext{2}{Department of Physics and Astronomy, University of Nevada Las Vegas, NV 89154, USA; zhang@physics.unlv.edu}

\begin{abstract}

The turbulence in the diffuse intergalactic medium (IGM) plays an important role in various astrophysical processes across cosmic time, but 
it is very challenging to constrain its statistical properties both observationally and numerically.  
Via the statistical analysis of turbulence along different sightlines toward a population of fast radio bursts (FRBs), 
we demonstrate that FRBs provide a unique tool to probe the intergalactic turbulence.
We measure the structure function (SF) of dispersion measures (DMs) of FRBs
to study the multi-scale electron density fluctuations 
induced by the intergalactic turbulence.
The SF has a large amplitude and a Kolmogorov power-law scaling with angular separations,
showing large and correlated DM fluctuations over a range of length scales. 
Given that the DMs of FRBs are IGM dominated,
our result tentatively suggests that the intergalactic turbulence has a Kolmogorov power spectrum and an outer scale on the order of $100$ Mpc.

\end{abstract}

\section{Introduction}

Turbulence is ubiquitous in astrophysical plasmas in both local and high-redshift universe
\citep{BraL14}.
It accompanies the large scale structure formation 
and amplifies cosmic magnetic fields 
\citep{Ryu08}.
It influences multi-scale diverse astrophysical processes, such as star formation 
\citep{Mckee_Ostriker2007},
cosmic ray propagation 
\citep{XY13}, 
magnetic reconnection and particle acceleration
\citep{Zh11,LaR20}.

The fundamental problem of turbulence is turbulent statistics 
\citep{Chan49}.
The statistical studies of astrophysical turbulence greatly benefit from the 
recent development of turbulence measurement techniques, including, e.g., 
the principal component analysis
\citep{Hey97},
Velocity Channel Analysis
\citep{LP00},
Velocity Coordinate Spectrum 
\citep{LP06},
core velocity dispersion 
\citep{Qi12},
polarization variance analysis 
and polarization spatial analysis 
\citep{LP16},
velocity gradient technique
\citep{Yu17}.
Statistical measurements of velocity field 
\citep{Chep10, Xu20,Li20},
density field 
\citep{Armstrong95,Burk09,CheL10},
magnetic field 
\citep{Han04,Gae11},
and other observables associated with turbulence 
\citep{XZ16,XuZ17}
reveal both the properties and important roles of turbulence in 
the interstellar medium (ISM) and intracluster medium (ICM).

Turbulence in the intergalactic medium (IGM) is closely related to the formation of large scale structure in the universe.
For the turbulence of non-primordial origin, 
the possible driving mechanisms include cosmological shocks in filaments 
\citep{Ryu08}
and supernovae-driven galactic outflows 
\citep{Evo11}.
The intergalactic turbulence significantly affects 
the dynamics of baryon fluid, galaxy-IGM interplay, amplification of magnetic fields, and enrichment of metals in the IGM through cosmic time
\citep{Evo10}. 
Despite the observational and numerical evidence indicating the presence of intergalactic turbulence 
(e.g., \citealt{Rau01,Lap11}),
unlike the turbulence in the ISM and ICM, 
the statistical properties of intergalactic turbulence are poorly constrained by observations, as 
the detection and measurements of the tenuous IGM are very challenging. 
Moreover, the statistical analysis of the large-scale intergalactic turbulence is infeasible
with current computational resources 
\citep{Lap11}.

Transient extragalactic radio bursts, such as fast radio bursts (FRBs), 
have their dispersion measures (DMs) dominated by the contribution of the IGM
\citep{Lor07,Tho13,Pat16}
and are powerful probes of the intergalactic turbulence 
\citep{Macq13,XZ16,Rav16}. 
Besides the scattering effect that causes the temporal broadening for individual FRBs
\citep{Macq13,Zhu18},
density fluctuations induced by intergalactic turbulence can also give rise to fluctuations in DMs of different FRBs. 
Similar to using Galactic pulsars to sample the interstellar turbulence
\citep{Armstrong95,XuZ17}, 
we can also use a substantial population of FRBs to sample the intergalactic turbulence.
With a range of separations between sight lines through the IGM, 
FRBs can provide the measurement on the scale-dependent DM fluctuations induced by the multi-scale intergalactic turbulence. 
For the first time, we perform a statistical measurement of the intergalactic turbulence by using a population of FRBs. 
In this Letter, 
{we apply the statistical method developed by 
\citet{LP16} 
(hereafter LP16)
for extended sources to point sources. 
The same statistical approach can also be used for e.g., other extragalactic point sources 
\citep{XuZ16},
molecular cloud cores
\citep{Xu20}, 
Galactic pulsars, to study the fluctuations of observables in various media
and the associated astrophysical processes.}
The basic formalism of the statistical method is presented in \S 2. 
In \S 3, we compare the measured structure function of DMs of FRBs with our theoretical expectation. 
Discussion and conclusions are given in \S 4.

\section{Structure function analysis of DMs}
\label{sec: sfdm}

In a turbulent medium, 
we consider that the correlation function (CF) of electron density fluctuations $\delta n_e$ follows the 
power-law scaling, 
\begin{equation}\label{eq: fopwcf}
\begin{aligned}
    \xi(R,\Delta l) &= \langle \delta n_e(\bm{X_1},l_1) \delta n_e (\bm{X_2},l_2)\rangle \\ 
            &=  \langle \delta n_e^2 \rangle   \frac{L_i^m}{L_i^m  +  (R^2 + \Delta l^2)^\frac{m}{2}} ,
\end{aligned}
\end{equation}
where $\bm{X}$ is the 2D position of the source on the sky plane, $l$ is the distance along the line of sight (LOS),
$R = |\bm{X_1} - \bm{X_2}|$ is the projected separation between sources,
$\Delta l = l_1 - l_2$, 
and the angle brackets denote an ensemble average.
$R$ can be converted to the angular separation $\theta$ by $\theta = R / L$.
Here $L$ is the size of the turbulent medium that extends from the observer to a distance $L$.
The above power-law form of CF is commonly used for describing fluctuations in observables induced by turbulence 
(LP16; \citealt{XuZ16,Xu20}).
The correlation length $L_i$ and the power-law index $m$ characterize the statistical properties of turbulence. 
$m$ is related to the 3D power-law index of a turbulent spectrum $\alpha$ by 
\begin{equation}
     \alpha = -m-3. 
\end{equation}
We note that for Kolmogorov turbulence, $m=2/3$ and $\alpha = -11/3$.

\begin{figure}[htbp]
\centering   
\subfigure[Case (1)]{
   \includegraphics[width=4cm]{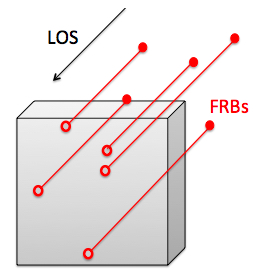}\label{fig: sketa}}
\subfigure[Case (2)]{
   \includegraphics[width=4cm]{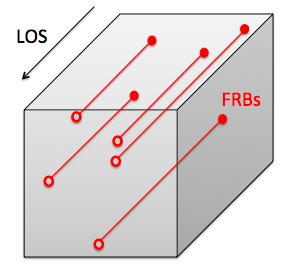}\label{fig: sketb}}
\caption{ Sketches of (a) a thin turbulent screen between the sources (FRBs) and the observer and 
(b) a turbulent volume along the entire LOS containing both the sources and the observer. 
The open circles indicate the 2D positions of FRBs projected on the sky plane. } 
\label{fig: sket}
\end{figure}

To calculate the structure function (SF) of dispersion measures
$\text{DM} = \int n_e dl $, 
where $n_e$ is the electron density, 
we consider two cases with 
(1) a single thin turbulent screen between the sources and the observer with the screen thickness much smaller than 
the distances of the sources from the observer
(Fig. \ref{fig: sketa}), 
and (2) a turbulent volume along the entire LOS containing both the sources and the observer (Fig. \ref{fig: sketb}). 
In the former case, only the components of DMs from the turbulent screen are correlated.

Case (1): a thin turbulent screen.~
In this case, the SF of DMs is 
\begin{equation}
\begin{aligned}
     D(R) &= \langle [\text{DM} (\bm{X_1}) - \text{DM} (\bm{X_2})]^2 \rangle \\
             & = \Big\langle \Big[ \int_0^L dl n_e (\bm{X_1},l) - \int_0^L dl n_e (\bm{X_2}, l) \Big]^2 \Big\rangle \\
             & = 4  \langle \delta n_e^2 \rangle \int_0^L d\Delta l (L-\Delta l)  \\
             & ~~~~~~ \Bigg[    \frac{L_i^m}{L_i^m  +  \Delta l^m} -    \frac{L_i^m}{L_i^m  +  (R^2 + \Delta l^2)^\frac{m}{2}}  \Bigg] ,
\end{aligned}
\end{equation}
where the expression in Eq. \eqref{eq: fopwcf} is used. 
When the thickness of the turbulent screen $L$ is larger than $L_i$, it has asymptotic scalings in different regimes 
(LP16), 
\begin{subnumcases}
     { D(R) \approx  \label{eq: drthc} }
       4  \langle \delta n_e^2 \rangle L_i^{-m} L R^{m+1}, ~~~~~~~~R<L_i,  \label{eq: steiner}\\    
       4  \langle \delta n_e^2 \rangle L_i^m L R^{-m+1} ,~~~~ L_i<R<L,\\     
       4  \langle \delta n_e^2 \rangle L_i^m L^{-m+2}, ~~~~~~~~~~~ R > L.
\end{subnumcases}
For a steep turbulent spectrum dominated by large-scale turbulent fluctuations 
\citep{LP04}
with $\alpha < -3$, e.g., Kolmogorov turbulence, 
$L_i$ is the outer scale of density fluctuations, and only 
Eq. \eqref{eq: steiner} is applicable. 
We then have 
\begin{subnumcases}
     { D(R) \approx   \label{eq: drsscst}}
       4  \langle \delta n_e^2 \rangle L_i^{-m} L R^{m+1}, ~~~~~~~~R<L_i,  \\   
       4  \langle \delta n_e^2 \rangle L_i L , ~~~~~~~~~~~~~~~~~~~~~~R>L_i.
\end{subnumcases}
The dependence on $R$ is seen when $R$ is in the inertial range of turbulence ($< L_i$). 
At $R > L_i$, DMs become uncorrelated, and $D(R)$ remains constant.

Case (2): a turbulent volume along the entire LOS.~
In a different case with both the sources and the observer within the same turbulent volume, 
the SF of DMs is 
\begin{align}
     D(R,l_1,l_2) & =  \langle [\text{DM} (\bm{X_1},l_1) - \text{DM} (\bm{X_2},l_2)]^2 \rangle  \nonumber\\
                           & = \Big\langle \Big[ \int_0^{l_1} dl n_e (\bm{X_1},l) - \int_0^{l_2} dl n_e (\bm{X_2}, l) \Big]^2 \Big\rangle \nonumber\\
                           & \approx 2 D^+(R,l_+) + \frac{1}{2} \Lambda (\Delta l)^2 . \label{eq: tdzzd}
\end{align}   
Compared with Case (1) with a localized thin turbulent screen, 
the LOS integral here is not limited by the screen thickness, 
but is taken over the entire path from the observer to the source. 
The difference between the distances of sources $\Delta l $ only enters the second term. 
The dependence on $R$ appears in the first term 
(LP16),
\begin{equation}
\begin{aligned}
     D^+(R,l_+) & = 2  \langle \delta n_e^2 \rangle \int_0^{l_+} d\Delta l (l_+ -\Delta l)  \\
     & ~~~~~~ \Bigg[    \frac{L_i^m}{L_i^m  +  \Delta l^m} -    \frac{L_i^m}{L_i^m  +  (R^2 + \Delta l^2)^\frac{m}{2}}  \Bigg] ,
\end{aligned}
\end{equation}
where $l_+ = (l_1 + l_2)/2$. 
If we consider distant sources from the observer with $l_+ > L_i $, then we can reach 
\begin{subnumcases}
     { D^+(R,l_+) \approx   }
       2  \langle \delta n_e^2 \rangle L_i^{-m} l_+ R^{m+1}, ~~~~~~R<L_i,\\    
       2  \langle \delta n_e^2 \rangle L_i^m l_+ R^{-m+1} , L_i<R<l_+,\\     
       2  \langle \delta n_e^2 \rangle L_i^m l_+^{-m+2}, ~~~~~~~~~~~ R > l_+,
\end{subnumcases}
which is similar to Eq. \eqref{eq: drthc} but $L$ is replaced by $l_+$. 

For the second term of $D(R,l_1,l_2)$ in Eq. \eqref{eq: tdzzd},
the coefficient $\Lambda$ (LP16) can be simplified to 
\begin{align}
     \Lambda &= \xi(0,l_+) - \xi(R,l_+) + 2\xi(R,0) \nonumber\\
                       &= \langle \delta n_e^2 \rangle  \Bigg[ \frac{L_i^m}{L_i^m  +    l_+^m}  - 
                              \frac{L_i^m}{L_i^m  +  (R^2 + l_+^2)^\frac{m}{2}} \nonumber\\
                       & ~~~~   + 2   \frac{L_i^m}{L_i^m  +  R^m } \Bigg ], \nonumber\\
&  \approx
 \begin{cases} 
      2 \langle \delta n_e^2 \rangle     ,   ~~~ R< L_i  \\
      0,  ~~~~~~~~~~~~ R > L_i ,
 \end{cases}
\end{align}
where the expression in Eq. \eqref{eq: fopwcf} is used. 
We again consider a steep turbulent spectrum with $\alpha < -3$. Based on the above expressions, we now approximately {have} 
\begin{subnumcases}
     { D(R,l_1,l_2) \approx  \label{eq: extdsf} }
       4  \langle \delta n_e^2 \rangle L_i^{-m}  R^{m+1} l_+ +  \langle \delta n_e^2 \rangle (\Delta l)^2, \nonumber \\ 
                 ~~~~~~~~~~~~~~~~~~~~~~~~~~~~~~~~~~~~~~~~R<L_i,\\    
       4  \langle \delta n_e^2 \rangle L_i l_++  \langle \delta n_e^2 \rangle (\Delta l)^2, \nonumber \\ 
                 ~~~~~~~~~~~~~~~~~~~~~~~~~~~~~~~~~~~~~~~~R > L_i .
\end{subnumcases}
The quantities related to the distances of sources, i.e., $l_+$, $\Delta l$, do not distort the power-law scaling of $D(R,l_1,l_2)$
with $R$.

Next by averaging over $l_+$ and $\Delta l$, we can {obtain}
\begin{align}
    D(R) 
            & = \frac{1}{2L} \int_{-L}^L    \frac{d \Delta l}{L-\Delta l}  \int_{|\Delta l| /2 }^{L-|\Delta l| /2} d l_+  D(R,l_1,l_2) \label{eq: fadazf}\\
            & \approx
 \begin{cases} 
      2 \langle \delta n_e^2 \rangle L_i^{-m} L R^{m+1}  +\frac{1}{3} \langle \delta n_e^2 \rangle  L^2   ,   ~~~ R< L_i  \label{eq: resavfr}\\
      2 \langle \delta n_e^2 \rangle L_i L   +\frac{1}{3} \langle \delta n_e^2 \rangle  L^2  ,  ~~~~~~~~~~~~~~~~~ R > L_i .
 \end{cases}           
\end{align}
It has a similar form as Eq. \eqref{eq: drsscst}, but here 
$L$ is the length of the entire turbulent volume along the LOS containing both the sources and the observer. 
Besides, the extra second term at $R < L_i$ arises from the different distances of sources, 
which adds ``noise" to the scaling of $D(R)$ with $R$
revealed by the first term. 
In Eq. \eqref{eq: fadazf}, 
we assume that the distance differences can range from $0$ to $L$, but in fact for distant sources from the observer under consideration, 
they mainly occupy a subvolume
within the range of distances $[L_0, L]$, where $L_0 > L_i$.
Therefore $D(R)$ should be adjusted {as} 
\begin{align}
     D(R) &=  \frac{1}{2(L-L_0)} \int_{-L+L_0}^{L-L_0}    d \Delta l \\
   & ~~~~~  \frac{1}{L-L_0 -\Delta l}  \int_{L_0+|\Delta l| /2 }^{L-|\Delta l| /2} d l_+  D(R,l_1,l_2) \\
   & \approx 
    \begin{cases} 
      2 \langle \delta n_e^2 \rangle L_i^{-m} (L+L_0) R^{m+1}  +\frac{1}{3} \langle \delta n_e^2 \rangle  (L-L_0)^2   ,   \\
        ~~~~~~~~~~~~~~~~~~~~~~~~~~~~~~~~~~~~~~~~~~~~~~~~~~~~~~~~~~~~~~~~~~ R< L_i  \label{eq: disingd}\\
      2 \langle \delta n_e^2 \rangle L_i (L+L_0) +\frac{1}{3} \langle \delta n_e^2 \rangle  (L-L_0)^2 ,  \\
      ~~~~~~~~~~~~~~~~~~~~~~~~~~~~~~~~~~~~~~~~~~~~~~~~~~~~~~~~~~~~~~~~~ R > L_i .
 \end{cases}  
\end{align}
As a result, compared with Eq. \eqref{eq: resavfr}, we see an
increase of the first term by a factor of $(1+ L_0/L)$ and 
a decrease of the second term by a factor of $(L-L_0)^2 / L^2$,
leading to a significantly reduced level of ``noise".

\section{SF of DMs of FRBs}

\begin{figure}[htbp]
\centering   
   \includegraphics[width=9.5cm]{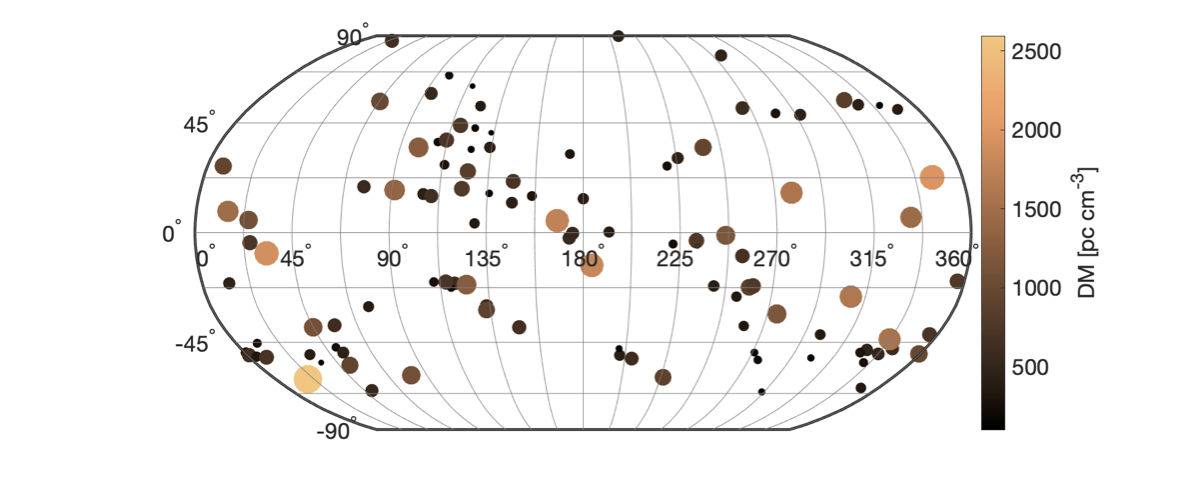}
\caption{ FRBs with measured DMs on the sky in Galactic coordinates. 
The circle size scales with DM. 
The color coding gives the DM values.  } 
\label{fig: map}
\end{figure}

\begin{figure*}[htbp]
\centering   
\subfigure[]{
   \includegraphics[width=8.5cm]{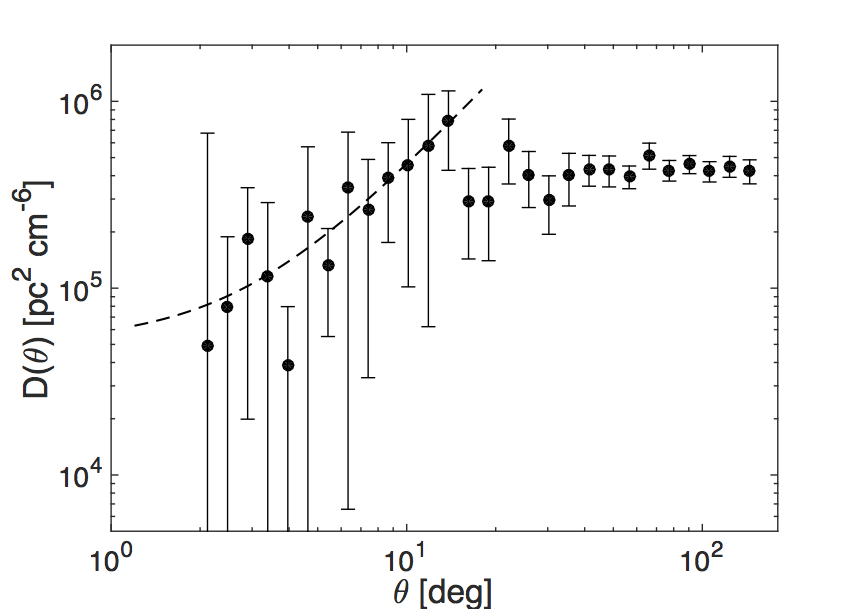}\label{fig: sf1}}
\subfigure[]{
   \includegraphics[width=8.5cm]{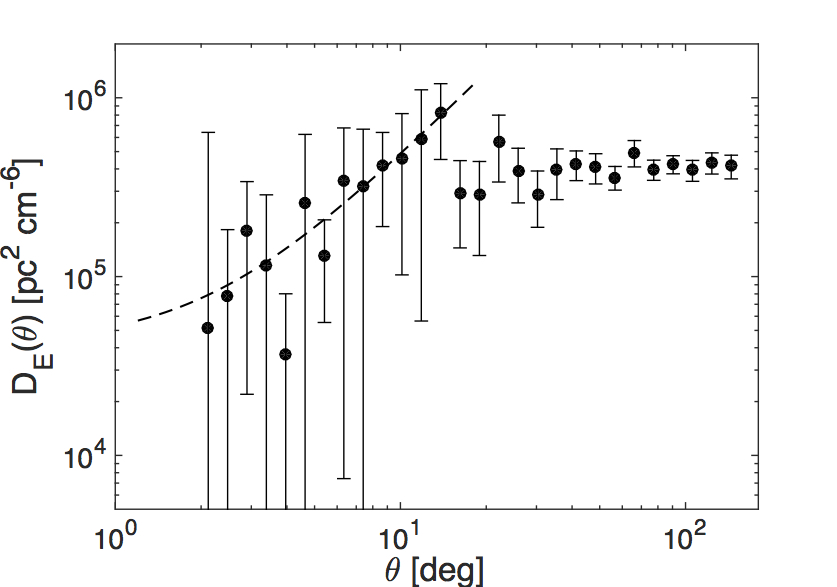}\label{fig: sfe}}
\caption{ (a) $D (\theta)$ vs. $\theta$ for 112 FRBs. Error bars indicate $95\%$ confidence intervals. 
The dashed line is the fit to the data points at small $\theta$ with the fitting function and parameters given by Eqs. \eqref{eq: fitfun} and \eqref{eq: fitpa}.
(b) Same as (a) but for $D_E (\theta)$. 
The dashed line shows the fit (Eq. \eqref{eq: fitfun}) with
$\alpha = 1.13\times10^4 \pm 1.31\times10^4$, 
$\beta = 1.60 \pm 0.43$, 
and $\gamma = 4.17\times10^4 \pm 6.50\times10^4$,
where the uncertainties are given at 68$\%$ confidence. } 
\label{fig: sf}
\end{figure*}

Using the most updated published population of FRBs 
\citep{Pat16}
\footnote{http://www.frbcat.org},
we calculate the SF of their measured total DMs as 
\begin{equation}
     D (\theta) = \langle (\text{DM} (\bm{X_1}) - \text{DM} (\bm{X_2}))^2 \rangle,
\end{equation}
which is the average value of the squared DM differences of all pairs of FRBs at a given angular separation. 
Here $\bm{X}$ is the projected position of an FRB on the sky plane, $\theta$ is the angular separation between projected positions,
and the angle brackets denote the spatial average at a fixed $\theta$. 
From the sky distribution of FRBs with measured DMs shown in Fig. \ref{fig: map}, 
we see that they sample the turbulent fluctuations along the LOS in different directions. 
So we are unlikely biased to detect the turbulent structure toward a particular direction. 
The result for the SF is displayed in Fig. \ref{fig: sf1}, where the error bars show $95\%$ confidence intervals. 
The error bars are larger toward a small $\theta$ due to the fewer number of pairs of FRBs available at a small $\theta$.
Based on the above analysis, we use a function 
\begin{equation}\label{eq: fitfun}
    D (\theta \lesssim 13.8^\circ)  [\text{pc$^2$ cm$^{-6}$}]  = \alpha  (\theta [^\circ] ) ^ \beta + \gamma  
\end{equation}
to fit the data points at small $\theta$.  
We find that for the best least-squares fit, there are 
\begin{equation}\label{eq: fitpa}
\begin{aligned}
  &    \alpha = 8595  \pm 1.03\times10^4, \\
  &     \beta = 1.68  \pm 0.44, \\
  &     \gamma = 5.13\times10^4  \pm 5.87\times10^4,
\end{aligned}
\end{equation}
where the uncertainties are given at $68\%$ confidence. 
$D(\theta)$ saturates and basically remains constant at $\theta > 13.8^\circ$.

The SF of DMs of FRBs at cosmological distances can be decomposed into its Galactic component $D_G$
and extragalactic component $D_E$, 
\begin{equation}\label{eq: gendedg}
\begin{aligned}
     &~~~~~~D(R) \\
     &= \langle [\text{DM}_G (\bm{X_1}) + \text{DM}_E(\bm{X_1})  - \text{DM}_G (\bm{X_2}) - \text{DM}_E (\bm{X_2})]^2 \rangle \\
             & = \underbrace{\langle [\text{DM}_G (\bm{X_1})  - \text{DM}_G (\bm{X_2}) ]^2 \rangle}_{D_G} \\
             & ~~~~ +  \underbrace{\langle [\text{DM}_E (\bm{X_1})  - \text{DM}_E (\bm{X_2}) ]^2 \rangle}_{D_E} , \\
\end{aligned}
\end{equation}
where $\text{DM}_G$ and $\text{DM}_E$ are the Galactic and extragalactic components of the total DM, respectively. 
We next consider two different cases with the power-law behavior of $D(\theta)$ dominated by 
(1) the Galactic ISM, 
or (2) the IGM.

(1) The Galactic ISM.~
If the DMs$_E$ toward different FRBs are uncorrelated, then $D_E$ is independent of $R$.
We can write Eq. \eqref{eq: gendedg} as 
\begin{equation}
    D(R) = D_G(R)  +  C,
\end{equation}
where $C$ is a constant representing $D_E$. 
In this situation, Case (1) in Section \ref{sec: sfdm} applies, and 
our Galaxy acts as a thin turbulent screen with the thickness $L$ given by the average path length through the Galactic ISM. 
We consider that the Galactic interstellar turbulence has 
a steep power-law spectrum  
and its driving scale is much smaller than $L$
\citep{Armstrong95,CheL10,Chep10}.
Accordingly, $D_G$ can be described by Eq. \eqref{eq: drsscst}, and thus there is 
\begin{subnumcases}
     { D(\theta) \approx   }
       4  \langle \delta n_e^2 \rangle L_i^{-m}  L^{m+2} \theta^{m+1} + C , \theta<L_i/L, \label{eq: galadmi}\\   
       4  \langle \delta n_e^2 \rangle L_i  L + C , ~~~~~~~~~~~~~~~~~~~~\theta>L_i/L,
\end{subnumcases}
where we use $\theta = R/L$ as the angular separation corresponding to $R$. 
We compare Eq. \eqref{eq: galadmi} with the fit to the measured $D(\theta)$ in Eq. \eqref{eq: fitfun}.
To explain the observations, there should be 
\begin{equation}
\begin{aligned}
      & m+1 = 1.68,  \\
      & 4  \langle \delta n_e^2 \rangle L_i^{-m}  L^{m+2} \Big(\frac{\pi}{180}\Big)^{m+1} = 8595, \\
      & C = 5.13\times10^4 , \\
      & \frac{L_i}{L} \approx 0.24.
\end{aligned}
\end{equation} 
From the above constraints one can easily get 
\begin{equation}
    \langle \delta n_e^2 \rangle L^2 [\text{pc$^2$ cm$^{-6}$}]  = 7.35\times10^5.
\end{equation}
It requires that the typical DM$_G$ of an FRB is 
\begin{equation}
    \text{DM}_G [\text{pc cm$^{-3}$}] \approx n_e L  \sim \sqrt{\langle \delta n_e^2 \rangle} L = 857.
\end{equation}
Obviously, this value is much larger than those of pulsars in the high Galactic latitude region where most FRBs were detected 
\citep{Cord19}.
In fact, the IGM is believed to be the dominant source of dispersion for most FRBs 
\citep{Iok03,Ino04,Lor07,Tho13}.
In Fig. \ref{fig: sfe}, we present the SF of DMs$_E$, where 
$\text{DM}_E = \text{DM} - \text{DM}_G$, and DM$_G$ is estimated based on the 
NE2001 Galactic electron density model
\citep{Cor02,Pat16}.
\footnote{Here we exclude the source with DM$_G > $ DM. }
The difference between Fig. \ref{fig: sf1} and Fig. \ref{fig: sfe} is marginal, which confirms the negligible Galactic contribution to 
$D(\theta)$.

(2) The IGM. ~
If the DMs$_E$ are correlated so that $D_E$ is a function of $R$, 
then $D(R)$ mainly reflects the statistical properties of the intergalactic turbulence given $D_E \gg D_G$.
By probing the intergalactic turbulence along the entire LOS, we are dealing with Case (2) in Section \ref{sec: sfdm}. 
Hence we approximately {have} 
\begin{align}
    D(\theta) & \approx D_E(\theta)  \\
   & \approx 
 \begin{cases} 
      2 \langle \delta n_e^2 \rangle L_i^{-m} (L+L_0) L^{m+1} \theta^{m+1}  \\
      ~~~~~~~~~~~~~~~~~~~+\frac{1}{3} \langle \delta n_e^2 \rangle  (L-L_0)^2   ,   
        ~~~~~~ \theta< L_i /L  \label{eq: frbing}\\
      2 \langle \delta n_e^2 \rangle L_i (L+L_0)  +\frac{1}{3} \langle \delta n_e^2 \rangle  (L-L_0)^2 ,  \\
      ~~~~~~~~~~~~~~~~~~~~~~~~~~~~~~~~~~~~~~~~~~~~~~~~~~~~~~~~~~~~ \theta > L_i /L,
 \end{cases}   
\end{align}
where $\theta = R/L$ and $L$ is the depth of the intergalactic turbulent volume that FRBs sample. 
Here we use Eq. \eqref{eq: disingd} under the consideration that FRBs are distant sources from the observer and the distances of most FRBs are 
larger than $L_0$, which can be constrained by the observational result (see below).

Similar to the earlier analysis, 
the comparison between Eq. \eqref{eq: frbing} and the fit to data (Eqs. \eqref{eq: fitfun} and \eqref{eq: fitpa}) leads to 
\begin{equation}
\begin{aligned}
    & m+1 = 1.68,  \\
    & 2 \langle \delta n_e^2 \rangle L_i^{-m} (L+L_0) L^{m+1} \Big(\frac{\pi}{180}\Big)^{m+1} = 8595 ,\\
    & \frac{1}{3} \langle \delta n_e^2 \rangle  (L-L_0)^2    = 5.13\times10^4, \\
    & \frac{L_i}{L} \approx 0.24.
\end{aligned}
\end{equation}
From these relations we obtain 
\begin{align}
    & m = 0.68,  \label{eq: turpinx}\\
    & \frac{L_0}{L} \approx 0.59,\\
    & \frac{L_i}{L} \approx 0.24.  \label{eq: corrtur}
\end{align}
Eq. \eqref{eq: turpinx} indicates that the intergalactic turbulence follows the Kolmogorov scaling ($m=2/3$). 
We note that the Kolmogorov scaling also applies to magnetized turbulence
\citep{GS95,LV99,CLV_incomp}, 
which would not be distorted by the presence of intergalactic magnetic fields 
\citep{Ryu08}.

\begin{figure*}[htbp]
\centering   
\subfigure[]{
   \includegraphics[width=8.5cm]{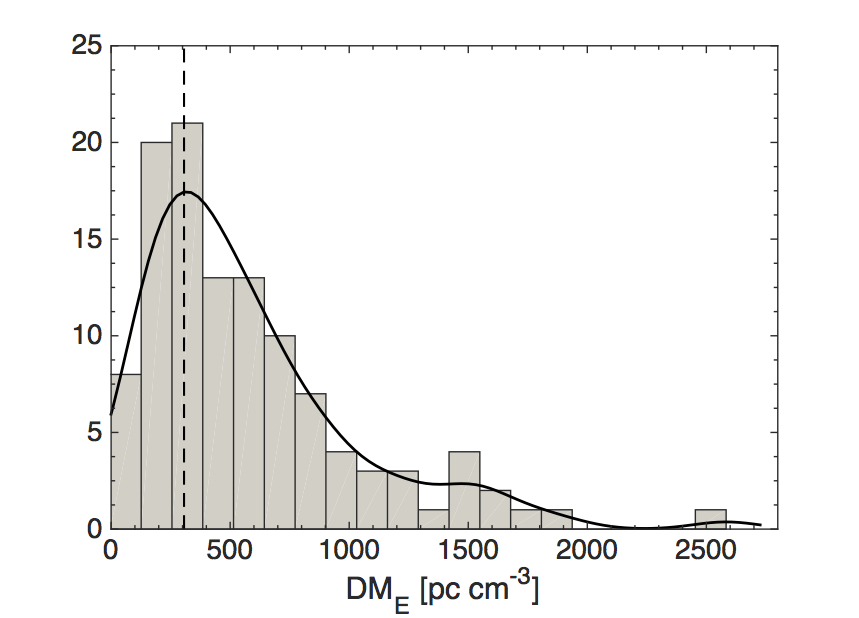}\label{fig: pdf}}
\subfigure[]{
   \includegraphics[width=8.5cm]{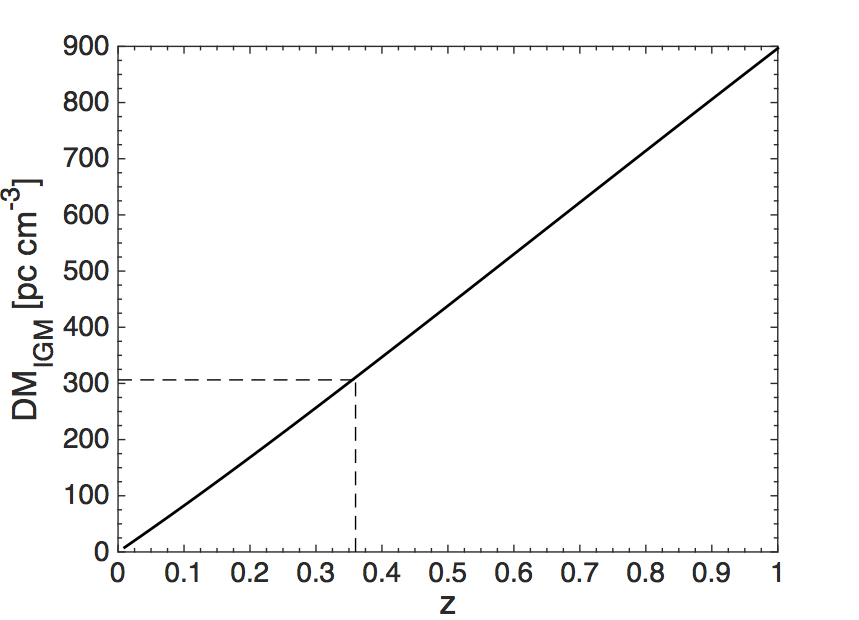}\label{fig: zdm}}
\caption{ (a) $\text{DM}_E$ distribution for the entire FRB sample. The thick solid line shows the 
kernel density estimate of the distribution. The peak of the distribution at $\text{DM}_{Ep} = 306.3$ pc cm$^{-3}$
is indicated by the vertical dashed line. 
(b) DM$_\text{IGM}$-$z$ relation (solid line). 
The dashed line marks $z \approx 0.36$ corresponding to $\text{DM}_{Ep}$.}
\end{figure*}

By using Eq. \eqref{eq: corrtur}, we can also evaluate the driving scale of intergalactic turbulence, 
which is about one order of magnitude smaller than $L$.
From $\text{DM}_E$ distribution (see Fig. \ref{fig: pdf}), where we subtract DM$_G$ based on the NE2001 model (see above), 
we find the peak at $\text{DM}_{Ep} \approx 306.3$ pc cm$^{-3}$.
The relation between the intergalactic component of DM, DM$_\text{IGM}$, and redshift $z$ was derived by 
\citet{Deng14}. 
Its numerical value 
\citep{Zha18}
\begin{equation}
     \text{DM}_\text{IGM} \approx 807~ \text{pc cm}^{-3} \int_0^z \frac{(1+z) dz}{ [\Omega_m (1+z)^3 + \Omega_\Lambda]^\frac{1}{2}}
\end{equation}
is shown in Fig. \ref{fig: zdm},
where $\Omega_m = 0.3089\pm 0.0062$ 
and $\Omega_\Lambda = 0.6911\pm0.0062$ are the matter density parameter and dark energy density parameter
\citep{Pla16}.
By assuming $\text{DM}_E \approx \text{DM}_\text{IGM} $, we see that 
the redshift corresponding to $\text{DM}_{Ep}$ is approximately $0.36$. 
The LOS comoving distance for $z = 0.36$ is 
$1455$ Mpc.
We adopt $L = 1455$ Mpc as the size of the intergalactic turbulent volume sampled by most FRBs
and obtain 
$L_i \approx 350$ Mpc 
(Eq. \eqref{eq: corrtur}) as the estimated driving scale of intergalactic turbulence. 
This is of the same order of magnitude as the scale of galaxy superclusters
\citep{Oort83},
indicative of a possible connection between the formation of superclusters and intergalactic turbulence.

{Our result can be treated as a tentative evidence for the Kolmogorov intergalactic turbulence up to the scale 
of the order of $100$ Mpc. 
Upcoming observations of a larger population of FRBs will be used for further testing the result. }

\section{Conclusions}

\begin{figure}[h!]
\centering   
   \includegraphics[width=8.5cm]{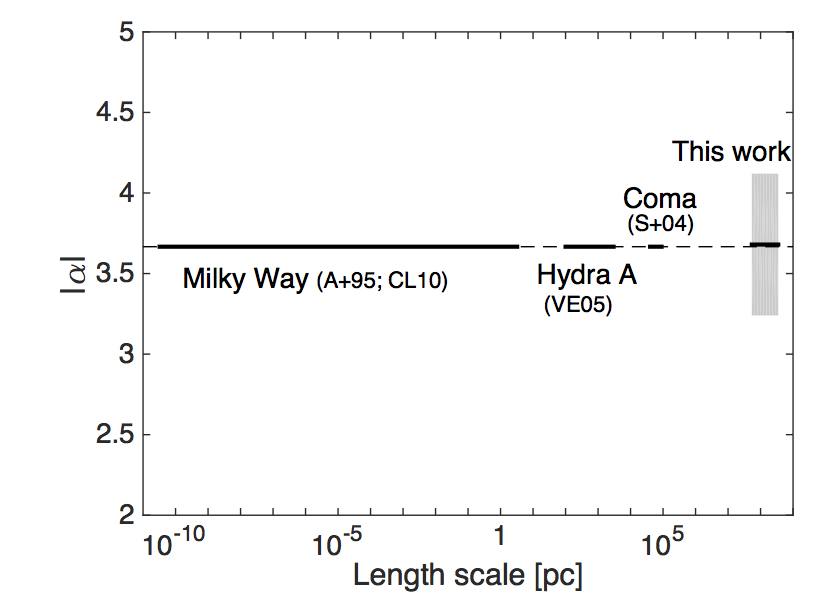}
\caption{ 3D power-law index $|\alpha|$ of turbulence vs. the range of length scales where the turbulent power spectrum is measured
in the Milky Way 
\citep{Armstrong95,CheL10},
Hydra A galaxy cluster
\citep{Vog05},
the Coma galaxy cluster
\citep{Schue04},
and in the IGM taken from this work. 
The shaded region indicates the uncertainty. 
The dashed line marks the Kolmogorov index. } 
\label{fig: tursky}
\end{figure}

Despite its astrophysical and cosmological significance,
the large-scale intergalactic turbulence and its statistical properties are poorly constrained by both observations and simulations.  
FRBs, with their cosmological distances and isotropic sky distribution, can serve as unique probes 
of the intergalactic turbulence. 
This work further demonstrates the universality of turbulence in the universe 
and provides information on the turbulence properties in the range of length scales beyond that of earlier measurements
(see Fig. \ref{fig: tursky}).

The SF of DMs of FRBs provide a direct measurement of the multi-scale turbulent fluctuations in electron density in the turbulent volume 
that FRB signals travel through. 
As the FRB signal passes through its host galaxy, the IGM, and the Milky Way, its DM includes multiple components. 
The resulting SF of DMs also contains the Galactic and extragalactic components. 
The latter is mainly contributed by the IGM under the assumption of generally small host contributions to DMs
\citep{Sha18}.
The power-law behavior of SF at small angular separations 
is expected from the energy cascade of turbulence in the inertial range. 
As the turbulent fluctuations in different host galaxies are uncorrelated, 
this power-law feature of SF can only come from either the Galactic interstellar turbulence or the intergalactic turbulence. 
The SF saturates at large angular separations as the electron density fluctuations are uncorrelated on scales beyond the inertial range of turbulence.

It is well established and tested that the Galactic ISM is turbulent and the turbulence has a characteristic Kolmogorov power spectrum 
in the warm ionized medium 
\citep{Armstrong95,Han04,CheL10}.
By comparing the observationally measured SF with the theoretically modeled SF dominated by the Galactic ISM, 
although the Kolmogorov power-law scaling can be explained, 
the Galactic DMs are too small to account for the measured SF value. 
This is also confirmed by the minor difference between the SF of total DMs and that of extragalactic DMs with the Galactic contributions 
subtracted based on the NE2001 model.

The large amplitude and power-law behavior of SF 
lead to the conclusion that 
the large and correlated DM fluctuations originate from the IGM. 
The comparison with the measured SF 
suggests that the intergalactic turbulence has 
a Kolmogorov scaling 
and a large driving scale on the order of $100$ Mpc corresponding to the transition angular separation where the SF saturates. 
The Kolmogorov velocity spectrum of cosmological turbulence up to the scale of 
superclusters ($\sim 100$ Mpc), which is the largest scale of inhomogeneities in the universe, 
is suggested by some cosmological models 
(e.g., \citealt{Oze78}). 
{However, 
it is known that the structure formation models involving primordial cosmic turbulence face some observational difficulties
\citep{Gol93},
and the role of hydrodynamics beyond the scales of galaxy clusters remains an unsolved problem.}

The current measured SF especially at small angular separations 
suffers from small source statistics and thus has a large uncertainty. 
Future observational tests with a larger population of FRBs are necessary for further studying the intergalactic turbulence 
and its cosmological implications {on structure formation scenarios. }

\acknowledgments
S.X. acknowledges the support for Program number HST-HF2-51400.001-A provided by NASA through a grant from the Space Telescope Science Institute, which is operated by the Association of Universities for Research in Astronomy, Incorporated, under NASA contract NAS5-26555.

\bibliographystyle{apj.bst}

\end{document}